\documentclass[twocolumn]{aastex63}

\usepackage{graphicx}
\usepackage{advdate}
\usepackage{amsmath}
\usepackage{appendix}

\newcommand{\asec}{\hbox to 1pt{}\rlap{$^{\prime\prime}$}.\hbox to 2pt{}}
\newcommand{\amin}{\hbox to 1pt{}\rlap{$^{\prime}$}.\hbox to 1pt{}}
\newcommand{\adeg}{\hbox to 1pt{}\rlap{$^{\circ}$}.\hbox to 2pt{}}

\accepted{for publication in {\it The Astrophysical Journal Letters}}
\shortauthors{Lauer, Postman, Spencer et al.}
\shorttitle{An Anomalous Cosmic Optical Background}

\begin{document}

\title{Anomalous Flux in the Cosmic Optical Background Detected With New Horizons Observations}

\author{Tod R. Lauer}
\affil{NSF's National Optical Infrared Astronomy Research
Laboratory,\footnote{The NSF's OIR Lab is operated by AURA, Inc.
under cooperative agreement with NSF.}
P.O. Box 26732, Tucson, AZ 85726}

\author{Marc Postman}
\affil{Space Telescope Science Institute,\footnote{Operated by AURA, Inc., for the National Aeronautics and Space Administration}
3700 San Martin Drive, Baltimore, MD 21218}

\author{John R. Spencer}
\affil{Department of Space Studies, Southwest Research Institute, 1050 Walnut St., Suite 300, Boulder, CO 80302}

\author{Harold A. Weaver}
\affil{The Johns Hopkins University Applied Physics Laboratory,
Laurel, MD 20723-6099}

\author{S. Alan Stern}
\affil{Space Science and Engineering Division, Southwest Research Institute, 1050 Walnut St., Suite 300, Boulder, CO 80302}

\author{G. Randall Gladstone}
\affil{Southwest Research Institute, San Antonio, TX 78238}
\affil{University of Texas at San Antonio, San Antonio, TX 78249}

\author{Richard P. Binzel}
\affil{Department of Earth, Atmospheric, and Planetary Sciences, Massachusetts Institute of Technology, Cambridge, MA 02139}

\author{Daniel T. Britt}
\affil{Department of Physics, University of Central Florida, Orlando, FL 32816}

\author{Marc W. Buie}
\affil{Department of Space Studies, Southwest Research Institute, 1050 Walnut St., Suite 300, Boulder, CO 80302}

\author{Bonnie J. Buratti}
\affil{Jet Propulsion Laboratory, California Institute of Technology, Pasadena, CA 91109}

\author{Andrew F. Cheng}
\affil{The Johns Hopkins University Applied Physics Laboratory,
Laurel, MD 20723-6099}

\author{W.M. Grundy}
\affil{Lowell Observatory, Flagstaff, AZ 86001}

\author{Mihaly Hor\'{a}nyi}
\affil{Laboratory for Atmospheric and Space Physics, University of Colorado, Boulder, CO 80303}

\author{J.J. Kavelaars}
\affil{National Research Council of Canada, Victoria BC \& Department of Physics and Astronomy, University of Victoria, Victoria, BC}

\author{Ivan R. Linscott}
\affil{Independent consultant, Mountain View, CA 94043}

\author{Carey  M. Lisse}
\affil{The Johns Hopkins University Applied Physics Laboratory,
Laurel, MD 20723-6099}

\author{William B. McKinnon}
\affil{Dept. of Earth and Planetary Sciences and McDonnell Center for the Space Sciences, Washington University, St. Louis, MO 63130}

\author{Ralph L. McNutt}
\affil{The Johns Hopkins University Applied Physics Laboratory,
Laurel, MD 20723-6099}

\author{Jeffrey M. Moore}
\affil{NASA Ames Research Center, Space Science Division, Moffett Field, CA 94035}

\author{J. I. N\'{u}\~{n}ez}
\affil{The Johns Hopkins University Applied Physics Laboratory,
Laurel, MD 20723-6099}

\author{Catherine B. Olkin}
\affil{Department of Space Studies, Southwest Research Institute, 1050 Walnut St., Suite 300, Boulder, CO 80302}

\author{Joel W. Parker}
\affil{Department of Space Studies, Southwest Research Institute, 1050 Walnut St., Suite 300, Boulder, CO 80302}

\author{Simon B. Porter}
\affil{Department of Space Studies, Southwest Research Institute, 1050 Walnut St., Suite 300, Boulder, CO 80302}

\author{Dennis C. Reuter}
\affil{NASA Goddard Space Flight Center, Greenbelt, MD 20771}

\author{Stuart J. Robbins}
\affil{Department of Space Studies, Southwest Research Institute, 1050 Walnut St., Suite 300, Boulder, CO 80302}

\author{Paul M. Schenk}
\affil{Lunar and Planetary Institute, Houston, TX 77058}

\author{Mark R. Showalter}
\affil{SETI Institute, Mountain View, CA 94043}

\author{Kelsi N. Singer}
\affil{Department of Space Studies, Southwest Research Institute, 1050 Walnut St., Suite 300, Boulder, CO 80302}

\author{Anne. J. Verbiscer}
\affil{University of Virginia, Charlottesville, VA 22904}

\author{Leslie A. Young}
\affil{Department of Space Studies, Southwest Research Institute, 1050 Walnut St., Suite 300, Boulder, CO 80302}

\begin{abstract}

We used New Horizons LORRI images to measure the optical-band ($0.4\lesssim\lambda\lesssim0.9{\rm\mu m}$) sky brightness within a high galactic-latitude field selected to have reduced diffuse scattered light from the Milky Way galaxy (DGL), as inferred from the IRIS all-sky $100~\mu$m map. We also selected the field to significantly reduce the scattered light from bright stars (SSL) outside the LORRI field. Suppression of DGL and SSL reduced the large uncertainties in the background flux levels present in our earlier New Horizons COB results. The raw total sky level, measured when New Horizons was 51.3 AU from the Sun, is $24.22\pm0.80{\rm ~nW ~m^{-2} ~sr^{-1}}.$  Isolating the COB contribution to the raw total required subtracting scattered light from bright stars and galaxies, faint stars below the photometric detection-limit within the field, and the hydrogen plus ionized-helium two-photon continua. This yielded a highly significant detection of the COB at ${\rm 16.37\pm 1.47  ~nW ~m^{-2} ~sr^{-1}}$ at the LORRI pivot wavelength of 0.608 $\mu$m. {This result is in strong tension with the hypothesis that the COB only comprises the integrated light of external galaxies (IGL) presently known from deep HST counts. Subtraction of the estimated IGL flux from the total COB level leaves a flux component of unknown origin at ${\rm 8.06\pm1.92 ~nW ~m^{-2} ~sr^{-1}}.$ Its amplitude is equal to the IGL.}

\end{abstract}


\section{A Targeted Observation of the Cosmic Optical Background}

The cosmic optical background (COB) is the flux of visible light photons averaged over the surface of the observable Universe. As it integrates over all processes that generate optical-band photons, it is a test of how well we understand what that integral should comprise. One way to pose this question is to ask if the galaxies that we see in cosmologically deep surveys are sufficient to account for the COB, or if there are significant sources of light yet to be recognized \citep{cooray}.

NASA's New Horizons spacecraft, which is presently over 50 AU away from the Sun, is an excellent platform for COB observations. Its sky is completely free of zodiacal light (ZL), which is sunlight scattered by interplanetary dust. ZL strongly dominates the sky brightness in the inner solar system. \citet{zemcov} produced a ``proof of concept" demonstration that New Horizons' LORRI camera \citep{lorri, lorri2} should be useful for COB observations, but had to contend with the dearth of useful archival images available at the time for measuring the COB flux.

\citet{l21}, in contrast, had a rich set of deep images to draw from and conducted a thorough examination of the calibration of New Horizons' LORRI camera for low light-level observations.  Based on seven fields, they measured the COB flux to be in the range ${\rm 15.9\pm 4.2\ (1.8~stat., 3.7~sys.) ~nW ~m^{-2} ~sr^{-1}}$ to ${\rm 18.7\pm 3.8\ (1.8~stat., 3.3 ~sys.)~ nW ~m^{-2} ~sr^{-1}}$ at the LORRI pivot wavelength of 0.608 $\mu$m, where the range reflects two different DGL corrections (diffuse galaxy light from the Milky Way scattered by infrared cirrus). When the estimated integrated light of galaxies (IGL) fainter than the LORRI photometric detection-limit was subtracted from this flux, a component of unknown origin in the range ${\rm 8.8\pm4.9\ (1.8 ~stat., 4.5 ~sys.) ~nW ~m^{-2} ~sr^{-1}}$ to $ 11.9\pm4.6~{\rm(1.8 ~stat., 4.2 ~sys.) ~nW ~m^{-2} ~sr^{-1}}$ remained. These measures are the most significant detections of the COB, and any unknown non-IGL component, to date.

The \citet{l21} image sets, however, were still drawn from archival observations.  The strongest foreground sources of light were DGL and scattered starlight (SSL) from bright field stars entering the LORRI camera from large angles. DGL and SSL vary strongly over the sky, however, which means that fields can be targeted that greatly minimize the contributions of both foregrounds. In this work we selected a field for pointed New Horizons COB observations that was estimated to markedly reduce DGL and SSL, compared to even the darkest field in \citet{l21}. As our analysis builds on \citet{l21}, we will frequently refer the reader to that work (hereafter NH21) for brevity.

\section{Measuring the COB Flux}\label{sec:obs}

\subsection{Selecting the Sky Field}\label{sec:field}

To identify fields with low foregrounds, we computed the SSL and DGL intensity levels for 60,000 randomly distributed positions in a $7320$ deg$^2$ area of sky bounded by galactic latitude $|b| \ge 40^{\circ}$ and a requirement that the fields' solar elongation angles (SEA) were $> 90^{\circ}$. Measurement of the background sky levels in LORRI images as a function of SEA $<90^\circ$ shows that the camera accepts scattered sunlight from large angles, which means that scattered {\it starlight} must be accounted for in fields within the New Horizons shadow (${\rm SEA}>90^\circ$). Only fields with ${\rm SEA}>90^\circ$ are suitable for COB observations in order to avoid sunlight entering the camera.

We estimated the DGL component at each position from the strength of the $100~{\rm\mu m}$ flux, which is due to the thermal emission of infrared ``cirrus".  The fluxes are provided by the ``IRIS" reprocessing of the IRAS full-sky thermal-IR maps \citep{iris} As we discuss in NH21, we subtracted a constant {cosmic infrared background} (CIB) level of 0.78 MJy sr$^{-1}$ \citep{cib1, cib2} from the IRIS map. {Even though there is no significant zodiacal light background at the distances from the Sun where the observations were obtained, we still must correct for any residual zodi signature in the IRIS data that remains even after the IRIS team applied its major zodi-subtraction. In NH21 we show that there is indeed a residual zodi-signature remaining in the IRIS flux values and we apply a smooth correction to the fluxes as a function of ecliptic latitude (see Figure 16 and Equation 8 in NH21) to remove this residual zodiacal light from the map.}  

The preliminary SSL at each location in the sky was estimated by convolving stars with $V < 11$ mag drawn from the Tycho2 star catalog \citep{tycho2}, and the Yale Bright Star catalog v5.0 \citep{YBSC5}, with the New Horizons scattered light response measured from preflight calibrations and inflight images. At each position we included stars up to $45^{\circ}$ away. 

We then sorted the fields based on their combined SSL and DGL intensities to identify  fields with significantly reduced DGL and SSL foregrounds, as compared to our earlier fields. We gave highest priority to fields with low DGL intensities. Once a final field was selected we recomputed the SSL by adding in fainter stars ($11 \le V < 20$ mag) from the {\it Gaia} DR2 catalog \citep{gaia2016,gaia2018}.

The selected field center is at J2000 $\alpha=0\adeg0756,$ $\delta=-21\adeg5451;$ the galactic latitude is $b=-77\adeg1,$ and the ecliptic latitude is $\beta=-19\adeg7.$ This position has ${\rm SEA}=113\adeg9,$ putting the aperture of LORRI safely within the spacecraft shadow.  

At the time of the observations New Horizons was 51.3 AU from the Sun, thus no ZL foreground was present. However, the ecliptic latitude is still important for understanding the $100~{\rm\mu m}$ flux derived from the {\it Earth-based} IRIS maps needed to estimate DGL. The $100~{\rm\mu m}$ flux measured from the IRIS map at this position{ prior to any background subtractions or corrections,} is ${\rm 1.756\pm0.042 ~MJy ~sr^{-1}}.$ This value is the mean IRIS flux within a circular area of radius $0\adeg2$ centered on the above position. This area corresponds to the circle that fully inscribes the LORRI FOV.  We subtract the ${\rm 0.78 ~MJy ~sr^{-1}}$ CIB flux, and the NH21 residual ZL correction of ${\rm 0.724 ~MJy ~sr^{-1}}$ at $\beta=19\adeg7$ {from the mean map value of ${\rm 1.756 ~MJy ~sr^{-1}}$}, leaving ${\rm 0.252\pm0.055 ~MJy ~sr^{-1}}$ as the estimated flux from any IR-cirrus in the field.  With the \citet{zemcov} scaling coefficient, this implies a DGL flux of only $2.22\pm1.00{\rm  ~nW ~m^{-2} ~sr^{-1},}$ with most of the error due to the large uncertainty in the coefficient. This DGL value is only 43\%\ of the lowest DGL intensity of the seven NH21 fields.
The field is also predicted to have to have an SSL foreground of $5.18\pm0.40{\rm  ~nW ~m^{-2} ~sr^{-1},}$ only 74\%\ of the lowest SSL of the seven NH21 fields.  

\subsection{Images of the Field}

\begin{figure}[htbp]
\centering
\includegraphics[keepaspectratio,width=3.4 in]{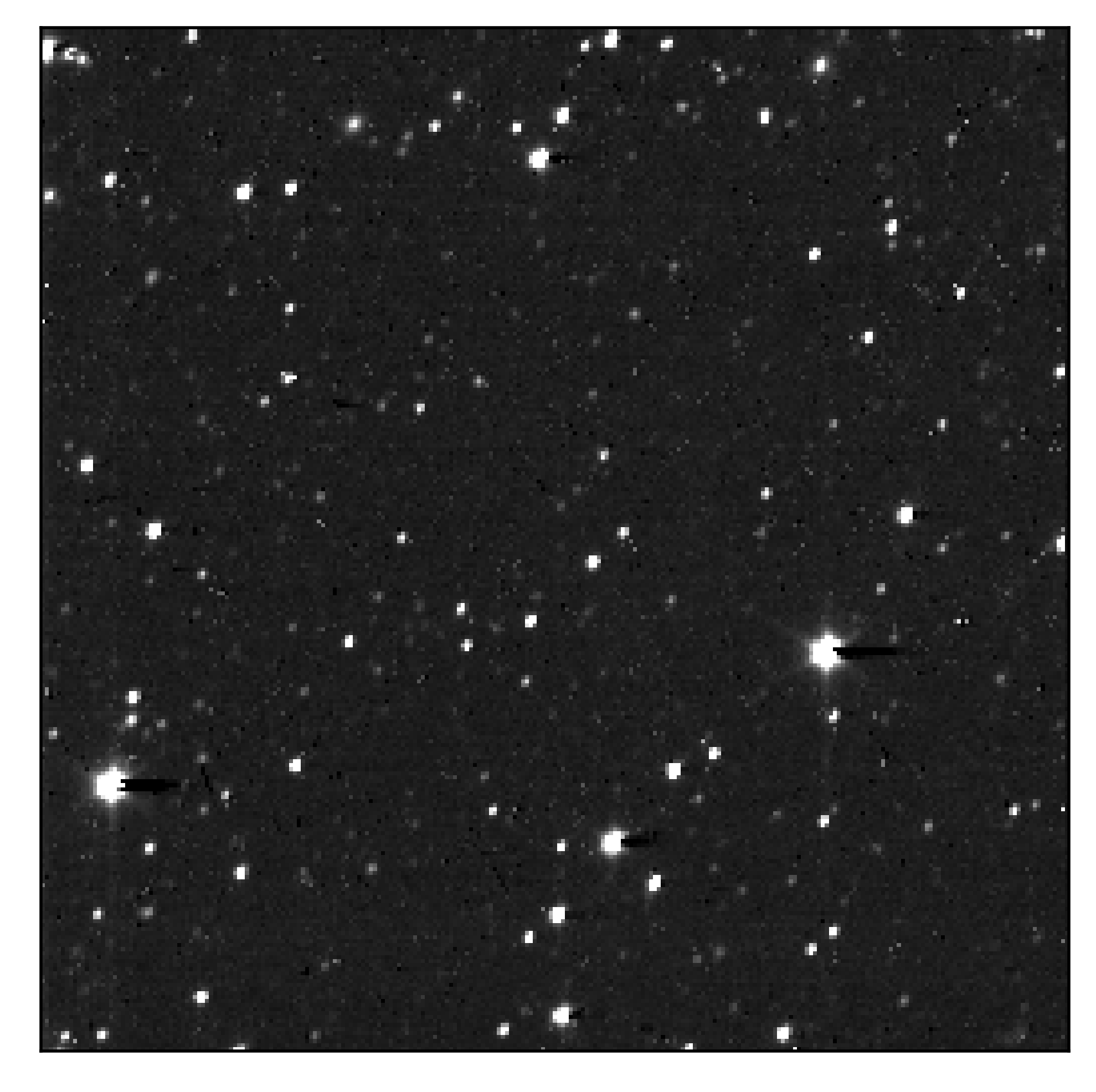}
\caption{An average of the first four images in the present dataset. The area is $17\amin4\times17\amin4.$ The display range is 50 DN (linear stretch starting at $-5$ DN). The faintest stars visible are at $V=19.1.$ The top of the field is at PA $139\adeg6.$}
\label{fig:image}
\end{figure}

The COB images were obtained with LORRI (the Long-Range Reconnaissance Imager) \citep{lorri, lorri2} on 2021 September 24 (UT) as a sequence of 16 65s exposures (only 30s exposures were used in NH21). The MET (mission elapsed time) IDs of the images were 0494832182 to 0494833607. The pointing was dithered by a few pixels between subsets of four images. A stack of the first subset is shown in Figure \ref{fig:image}. To avoid the LORRI ``background fade" anomaly associated with the activation of the camera (NH21), the exposure sequence was initiated five minutes after the camera was powered on. As a check, a fit to the sky levels of the 16 exposures as a function of time showed an insignificant drift of only $0.10\pm0.10$ DN (data number) over the 1040s duration of the sequence. 

In brief, LORRI is an unfiltered (white light) $1032\times1024$ pixel CCD imager mounted on a 20.9 cm aperture Cassegrain reflector. For deep observations, the camera is operated with $4\times4$ pixel binning, producing (raw) images in $257\times256$ pixel format, including a single bias/dark column. The pixel-scale in this mode is $4\asec08,$ which provides a $17\amin4$ field. LORRI's sensitivity extends from the blue ($0.4\mu{\rm m}$) to NIR ($0.9\mu{\rm m}$) and is defined by the CCD response and telescope optics.  The pivot wavelength is 0.608$~\mu$m.  The camera is operated with a gain of $19.4e^-$ per 1 DN, and the read-noise is $24e^-.$ In $4\times4$ mode the photometric zeropoint is $18.88\pm0.01$ AB magnitudes corresponding to a 1 DN/s exposure level \citep{lorri2}.

\subsection{Image Reduction}

The sky levels in the images are only slightly greater than 1 DN. The reduction of the images thus requires attention to a number of subtle effects that are only important at this level.  Rather than using calibrated (``Level 2'') images produced by the standard LORRI pipeline operated by the New Horizons project, we use the NH21 custom reduction of the raw (``Level 1") images to optimize accurate recovery of the faint sky signal. The first calibration step is to estimate the bias level by fitting a gaussian to the peak of the DN histogram of the bias column.  This provides bias values accurate to a fraction of a DN, while until recently, the standard pipeline selected the median integer DN level.

Subsequent to NH21, we discovered an error in the analogue to digital (A/D) conversion of the video signal produced by the LORRI CCD that required a small correction to be applied to the bias determination. Histograms of raw LORRI images showed that the measurement of the least-significant bit (LSB) during the A/D conversion was slightly in error, such that the set point of the LSB was 7\%\ too high, making even DN values 14\%\ more common than odd DN values (errors in the higher order bits were not evident). Analysis of the effects of this error were done following the precepts of \citet{wfpc1}, which discussed the diagnosis and correction of large A/D errors in the HST WFPC1 instrument. Briefly, bias values were recovered from simulated distributions of integer DN values generated from un-digitized gaussians of width appropriate to the LORRI readout noise.  Simulated A/D conversion was done with and without the LSB error, as the fractional location of the mean value of the distribution was  varied over a range of 1 DN. The measured mean value with the LSB error was always 0.02 DN too low, allowing for a simple additive correction to the measured bias levels.

The second step is to correct for the ``jail bar" pattern, where bias level of the even-numbered columns in the CCD are offset by $+0.5$ or $-0.5$ DN from that of the odd-numbered columns (which includes the bias column).  The sign of the offset is set randomly when the camera is powered on; in the present sequence the offset of the even columns is $+0.5$ DN.  This calibration step is not included in the standard pipeline.  The final calibration steps are subtraction of a ``super-bias" frame, charge-smear correction, and standard flat-field calibration.  The charge-smear correction is an improved version of that in the standard pipeline \citep{lorri2}, and we also exclude bright cosmic ray hits and negative amplifier under-shoot artifacts associated with over-exposed stars from the charge-smear calculations, as they are not smeared.

\subsection{Measuring the Sky Level}

The procedures for measuring the sky level are discussed extensively in NH21. In brief, we measure the sky for each individual exposure by first masking out foreground stars, galaxies, hot pixels, and cosmic ray events, and then fitting a gaussian to the peak of the intensity histogram of the remaining unmasked pixels. Masking is done by flagging all pixels above 8 DN intensity, and excluding all pixels within 3 pixels or $12''$ in radius around that pixel. This threshold is somewhat arbitrary; it is a compromise between detecting faint sources versus selecting on background noise. Low level wings at larger radii from  the stars do remain in the image, but these are corrected for in the estimation of the scattered starlight (SSL) components in field.  In practice the masking procedure deletes all objects with $V<19.9$ (this threshold is 0.8 mag deeper than the $V<19.1$ used in NH21, given the present 65s, rather than 30s exposures). { While nearly all the objects masked are stars, the galaxies deleted need to be accounted for, as their flux should be included in the COB. The LORRI angular resolution is too poor to allow classification of most galaxies above the detection threshold as non-stellar, thus our solution is to add the light from masked galaxies with $V<19.9,$ as catalogued by the PANSTARRS survey \citep{panstarrs}, to the IGL flux (see $\S$\ref{sec:igl}).}

\begin{figure}[htbp]
\centering
\includegraphics[keepaspectratio,width=3.5 in]{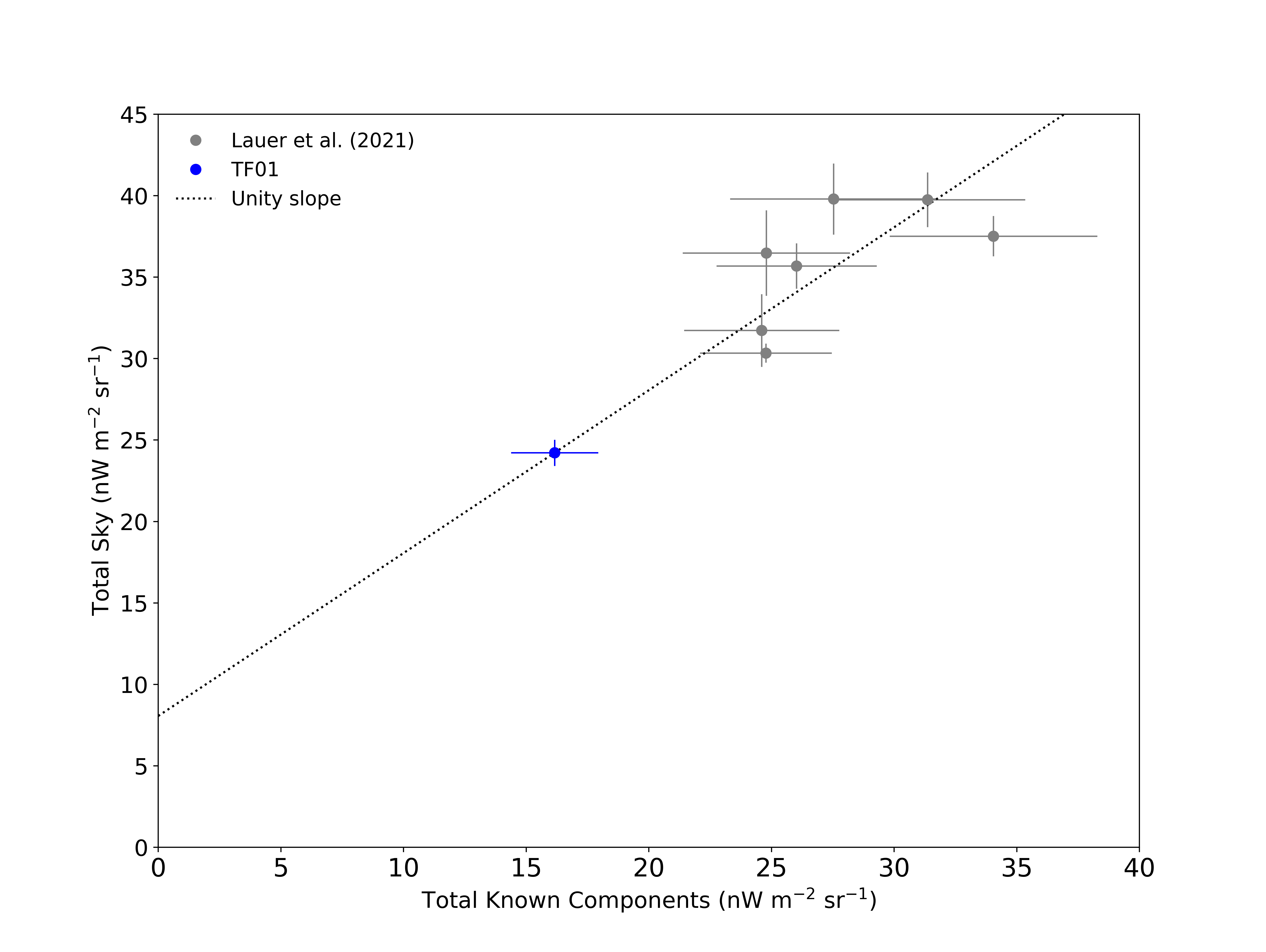}
\caption{The total sky levels for the present (TF01 = Test Field 1) and seven NH21 fields are plotted as function of the total known flux components present. A line with unit slope going through the point representing the present field is shown. This demonstrates that the total sky level in the present field decreased by the amount expected as compared to the NH21 fields, given its reduced foreground flux components.}
\label{fig:skycomp}
\end{figure}

The histogram fitting algorithm is designed to take into account fine scale structure of the distribution of pixel intensity values that results from the image calibration operations applied to the initially integer raw pixel values.  The histogram fitting procedure also ignores all pixels with values well away from the histogram peak.  In application we find the sky following the masking procedure is only 7\%\ less than the sky measured with no masking at all. The average sky value of the 16 images is $1.058\pm0.035$ DN, or a V-band surface brightness of 26.4 mag/arcsec$^2;$ { the associated error is statistical and is the error in the mean of the 16 images.} This corresponds to $24.22\pm0.80{\rm ~nW ~m^{-2} ~sr^{-1}}$ in flux units at the LORRI pivot wavelength of 0.608 $\mu$m. As shown in Figure \ref{fig:skycomp}, this sky level is significantly less than the typical raw sky levels of the 7 fields of NH21, but is essentially as expected given the estimated reduction of the DGL and SSL components.

\section{The Cosmic Optical Background Flux}

\begin{figure}[hbtp]
\centering
\includegraphics[keepaspectratio,width=3.5 in]{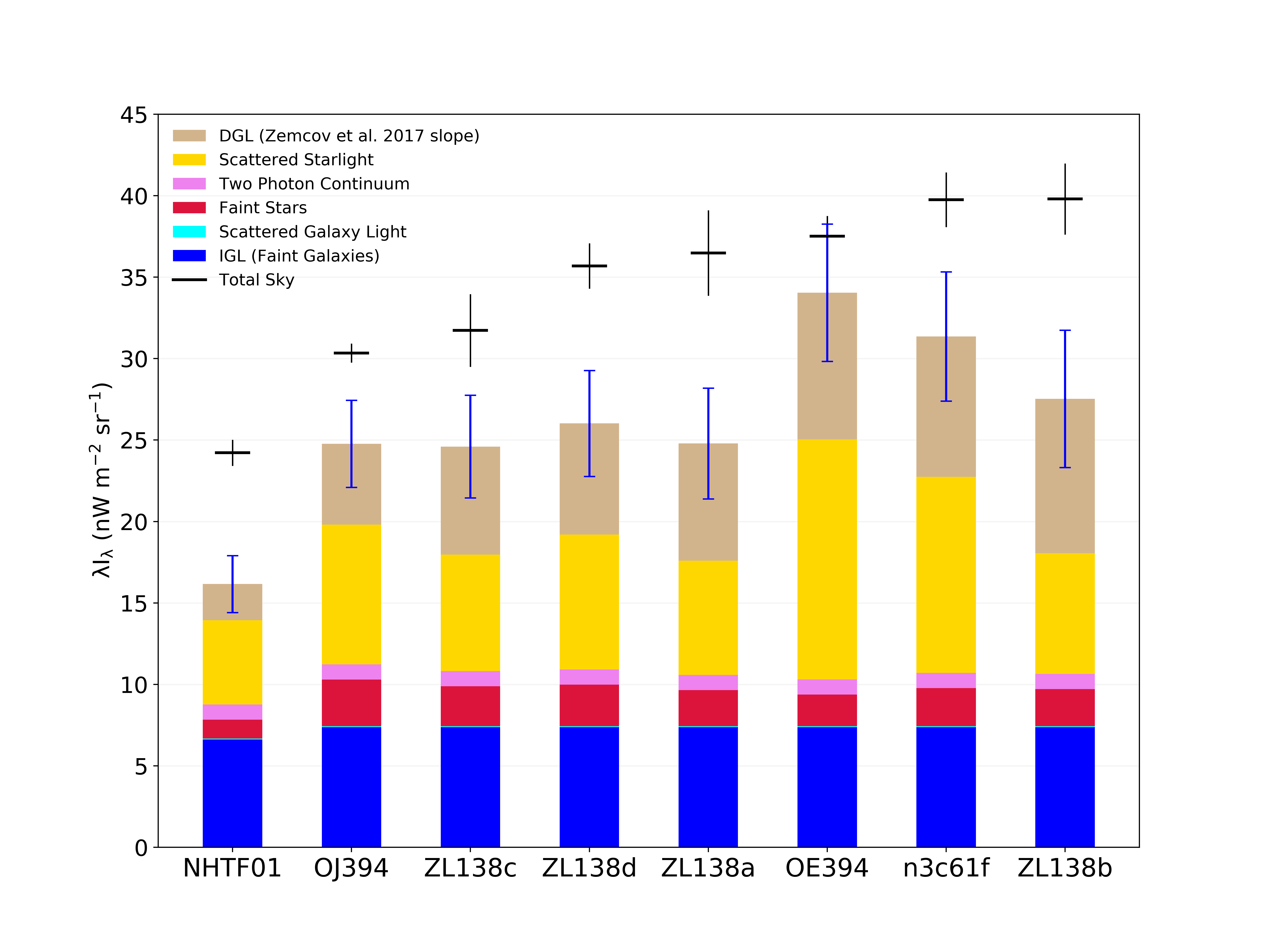}
\caption{A stacked bar chart showing the amplitudes of the known sky components for the present field (leftmost bar) as compared to the seven NH21 fields. The black horizontal lines with error-bars show our measured total sky values and their uncertainties for each field. {The small flux from bright galaxies masked out in the LORRI field is not included in this figure.}}
\label{fig:skybars}
\end{figure}

Isolating the COB flux from the total sky requires correcting for a number of foreground sources.  We describe these in detail in NH21, but present their specific contributions to the present field here. A summary of the decomposition of the total sky is shown in Figure \ref{fig:skybars}. {The fluxes in all components and their associated errors are listed in Table \ref{tab:decomp}. We break down the errors into systematic and statistical terms, as we discussed in detail in NH21.  Understanding which uncertainties are systematic is critical when combining the measurements in several fields as we did in NH21.  In the present case of a single field the errors in all flux components are independent, but again, this is no longer true when we compare the present results to those in NH21.}

\subsection{Scattered Light from Bright Stars (SSL) and Galaxies (SGL)}

As noted in $\S\ref{sec:field}$ the field was selected for its low SSL of $5.17\pm0.52{\rm  ~nW ~m^{-2} ~sr^{-1}}.$  The error {is systematic} and is dominated by uncertainty in the New Horizons scattered light function.  The SGL term comes from scattered light contributed by bright galaxies outside the LORRI field. The surface density of bright galaxies is so low that this flux, $0.07\pm0.01{\rm  ~nW ~m^{-2} ~sr^{-1}},$ is almost negligible.

\begin{deluxetable}{lccc}
\tabletypesize{\scriptsize}
\tablecolumns{3}
\tablewidth{0pt}
\tablecaption{Sky Flux Decomposition}
\tablehead{\colhead{Component}&\colhead{${\rm ~nW~m^{-2}~sr^{-1}}$}&Stat.&Sys.}
\startdata
{\bf Total Sky} &$24.22\pm0.80$&0.80&0.00\\
$-~$Scattered Starlight (SSL) &$\phantom{0}5.17\pm0.52$&0.00&0.52\\
$-~$Scattered Milky Way Light (DGL) &$\phantom{0}2.22\pm1.00$&0.32&0.95\\
$-~$Faint Stars (FSL) &\phantom{0}$1.16\pm0.18$&0.06&0.17\\
$-~$Two-photon continuum (2PC) &$\phantom{0}0.93\pm0.47$&0.00&0.47\\
$-~$Scattered Galaxy Light (SGL) &$\phantom{0}0.07\pm0.01$&0.00&0.01\\
$+\hskip 2pt$Bright Field Galaxies&$\phantom{0}1.70\pm0.07$&0.04&0.06\\
\hline
{\bf $=\hskip 2pt$Cosmic Optical Background}&$16.37\pm1.47$&0.86&1.19\\ 
$-~$Integrated Galaxy Light (IGL) &$\phantom{0}8.31\pm1.24$&0.78&0.97\\
\hline
{\bf $=\hskip 2pt$Anomalous Flux}&$\phantom{0}8.06\pm1.92$&1.16&1.53\\
\enddata
\end{deluxetable}\label{tab:decomp}

\subsection{Diffuse Galactic Light (DGL)}

The selection of the field was done to minimize the DGL foreground, as discussed in $\S\ref{sec:field}.$ We repeat the estimated DGL foreground flux here as $2.22\pm1.00{\rm  ~nW ~m^{-2} ~sr^{-1},}$ based on the \citet{zemcov} conversion of the $100~\mu$m flux, { as given by NH21 eqn.\ (7) with $C_{100}=9.8\pm3.9 {\rm  ~nW ~m^{-2} ~sr^{-1}.}$}

{As one check on our conversion we integrated the \citet{onishi} DGL coefficients derived from their WD01 and WLJ15 dust models (normalized to the $1.1 ~\mu{\rm m}$ measurement of their MBM32 field), as a function of wavelength over the LORRI response-function. This produced a mean coefficient only 10\% larger than ours (when we compute our coefficient for the same galactic latitude as their MBM32 field), which is well within our assumed $\sim40\%$ errors. 

As second check, we subtracted all the known flux components from the present and NH21 total sky fluxes, {\it except any estimate for the DGL flux}, and fitted a line to the residuals (which also contained the presumably-constant anomalous flux) as a function of $100~\mu{\rm m}$ flux. The slope of the line provides an estimate of the conversion coefficient. We recovered $C_{100}=10.1\pm5.2 {\rm  ~nW ~m^{-2} ~sr^{-1}.}$ in good agreement with the scaling used in NH21.}

{The systematic component in the DGL error dominates and is mainly due to the large error in the flux-conversion coefficient, with a smaller contribution from the error in the $100~\mu$m flux.  The statistical error is due to uncertainty in the correction of the $100~\mu$m map for residual zodiacal light. See Table \ref{tab:decomp} for both components.}

\subsection{Integrated Faint Starlight (FSL)}

The integrated light of faint stars (FSL) below the LORRI photometric detection limit is another foreground source that must be accounted for. Our approach is to integrate {\tt TRILEGAL} models \citep{tril2005, trilv16} of the expected population of faint stars within our fields, following the procedures developed in NH21. The only difference is that for this field the bright limit of the integral (Eq. 3 in NH21) is $V=19.9.$  For our specific field we estimate the FSL component as $1.16\pm0.18{\rm  ~nW ~m^{-2} ~sr^{-1}}.$ { The systematic and statistical components in the error (Table \ref{tab:decomp}) are due to uncertainties in the {\tt TRILEGAL} models parameters, and estimated fluctuations in the star counts, respectively.}

\subsection{The Two-Photon Continuum (2PC)}

The existence of a full-sky diffuse Ly-$\alpha$ background from the Milky Way \citep[e.g.][]{gladstone} means that there is also likely to be an associated hydrogen two-photon continuum \citep{two}.  New Horizons far UV spectroscopic observations taken of our field with the Alice instrument indeed show the existence of this component. The continuum extends to all wavelengths to the red of Ly-$\alpha$ and thus will have some flux contribution within the LORRI passband.  The Alice spectra also appear to show the minor presence of the analogous continuum from singly-ionized He. Using the spectral form of the two-photon continuum given by \citet{nuss}, we find that the contribution of the H and He$+$ continua integrated over the LORRI passband to be $0.93\pm0.47{\rm  ~nW ~m^{-2} ~sr^{-1}},$ a minor contribution to the total sky level. { We treat the error as systematic as it is assumed to be the same for both the present and NH21 COB fields.}

\subsection{Foregrounds from the Spacecraft}

Measuring the COB requires that the spacecraft environment itself is dark and does not contribute significant foreground light.  In NH21 we demonstrated that the spacecraft shadow was sufficiently dark such that sunlight had no indirect path of significance into the LORRI aperture for the SEAs of the COB fields.  We also considered it unlikely that the exhaust of the thrusters that stabilized the NH spacecraft could generate ice crystals sufficient to scatter light into LORRI. Subsequent to NH21, we identified two more effects of potential concern, the production of Cherenkov radiation and fluorescence, induced by energetic particles penetrating the LORRI field-flattening lenses. We estimate the strength of these two sources in Appendix \ref{sec:cher}, concluding that they do not contribute significant foreground flux.
{
Related to this, as part of the analysis done in NH21, we measured the dark current of the LORRI CCD at $0.334\pm0.039$ DN in 65 s, which is well within the CCD manufacturer’s speciﬁed performance.  There is no evidence for any strongly increased dark current due to irradiation of the CCD over the duration of the mission. We also note that the CCD dark/bias column will also witness the average level of any charge deposited directly in the CCD by energetic particles during an exposure.}

\subsection{The Total Cosmic Optical Background}

The COB is the flux that remains after we remove the artifactual scattered light foregrounds of bright stars (SSL) and galaxies (SGL) contributed by sources {\it outside} the LORRI field, as well as the flux from faint stars (FSL) and diffuse Milky Way light scattered by IR-cirrus (DGL) {\it within} the field from the observed total sky level.  { As the COB should also reflect the integral flux from all external galaxies, we have also added in the light from the bright galaxies that were present in the LORRI field, but masked out in the measurement of the total sky level. This correction is small and is discussed in detail in the next section.} The COB flux is thus $16.37\pm1.47~({\rm 0.86~stat.,~1.19~sys.)} {\rm  ~nW ~m^{-2} ~sr^{-1}}.$ The error is the simple quadrature sum of all the errors associated with the first six components tabulated in Table \ref{tab:decomp}. As we discuss in NH21, most of these errors are systematic, thus combining the present results with, say, the seven fields in our previous paper requires careful treatment of the correlated errors between all fields. For a single field, however, the errors can be regarded as statistical.  The present field provides the most significant detection of the COB to date.

\subsection{Integrated Galaxy Light (IGL)}\label{sec:igl}

The COB does contain the integrated flux from all galaxies that fall within the LORRI field.  This IGL component compared to the COB flux tests how well we understand the overall optical flux generated by the Universe.  

{For this analysis, the IGL is estimated in two steps: the bright IGL for galaxies with $V < 19.9$ that were masked during the sky estimation process and the faint IGL for galaxies below this LORRI detection threshold. The IGL for the bright galaxies ($V < 19.9$) is estimated by extracting non-stellar objects in our LORRI field of view from the second release of the PanSTARRS catalog available via the MAST archive \citep{panstarrs}. PanSTARRS objects with a difference greater than 0.05 mag between their PSF magnitude and their Kron magnitude in the PanSTARRS i-band are classified as galaxies. We compute a V-magnitude for each object from their g-band and r-band magnitudes provided by PanSTARRS. The transformation to V-mag from the g, r bands is derived from 8 templates of galaxy spectral energy distributions spanning the morphologies E, S0, Sa, Sb, Sc, and Ir types. We weight the templates by the morphological fractions observed in the field population of galaxies and derive an average $(V-g)$ vs. $(r-g)$ relationship over the redshift range $0 < z \le 1$, typical for brighter galaxies. We then derive the IGL flux contribution based on the $V$ magnitude and sum up the contributions for all PanSTARRS galaxies with $V < 19.9$ in the LORRI field of view. The IGL contribution computed in this way comes to $1.70\pm0.07\ (0.06~({\rm sys}),~0.04~({\rm stat})) {\rm ~nW ~m^{-2} ~sr^{-1}}$. The statistical error is derived from the photometric errors given in the PanSTARRS catalog and the systematic error is estimated by using different fitting functions and different SED templates for the $(V-g)$ vs. $(r-g)$ transformation.}

The precepts for estimating the faint IGL { due to galaxies at or below the $V=19.9$ detection threshold} are discussed at length in NH21. The faint IGL contribution in the present field is slightly reduced from that in our earlier fields due to the fainter $V=19.9$ bright limit to the galaxy flux integral (Eq. 3 in NH21). Our NH21 estimate for the uncertainty in the { faint} IGL of 30\%\ used in NH21 was conservative and was based on rough estimates of the variation in the faint end slope of the galaxy number count relations. 
{ We perform a more rigorous estimate of the uncertainty in our faint IGL flux by assessing the specific contribution to the error from the systematic terms (errors in the fits to the galaxy number counts) and from the statistical errors (cosmic variance). The two systematic errors associated with the fits to the galaxy number counts are from the errors in the coefficients to the power law fits used in NH21 and the error associated with the form of the fitting function (e.g., 4 power-laws vs a quadratic fit). The formal errors in the power law coefficients yield a fractional error of 13.1\% in the IGL flux. The difference between the IGL derived from the power law fits versus that derived using a quadratic fit to the galaxy counts yields a fractional change in the IGL of 6.6\%. Summing these two error components in quadrature yields a combined systematic fractional error of 14.7\% in the IGL flux.
The total error in the IGL must also include the statistical uncertainty due to the effects of cosmic variance over a single LORRI FOV. The cosmic variance error for a single LORRI field-of-view used in this work is the same as the single-field CV error adopted in NH21 \citep{cosvar2008} - which translates to an IGL fractional error of 11.8\%. Summing, in quadrature, this statistical error with the above systematic error yields a total fractional error of 18.8\% in the faint IGL flux. This is smaller than our conservative estimate of 30\% used in NH21 but represents a more accurate assessment of the error in the faint IGL component.
Our computed faint IGL flux in the current field is $6.61\pm1.24\ (0.97~({\rm sys}),~0.78~({\rm stat})) {\rm ~nW ~m^{-2} ~sr^{-1}}$.}

{Combining the bright and faint galaxy contributions to the IGL gives a total IGL flux for our field of $8.31\pm1.24\ (0.97~({\rm sys}),~0.78~({\rm stat})) {\rm ~nW ~m^{-2} ~sr^{-1}}$. This IGL corresponds to the expected light in the LORRI bandpass from all galaxies brighter than $V=30$ mag.}

\subsection{The Detection of a Significant Anomalous Flux Background}

\begin{figure}[htbp]
\centering
\includegraphics[keepaspectratio,width=3.5 in]{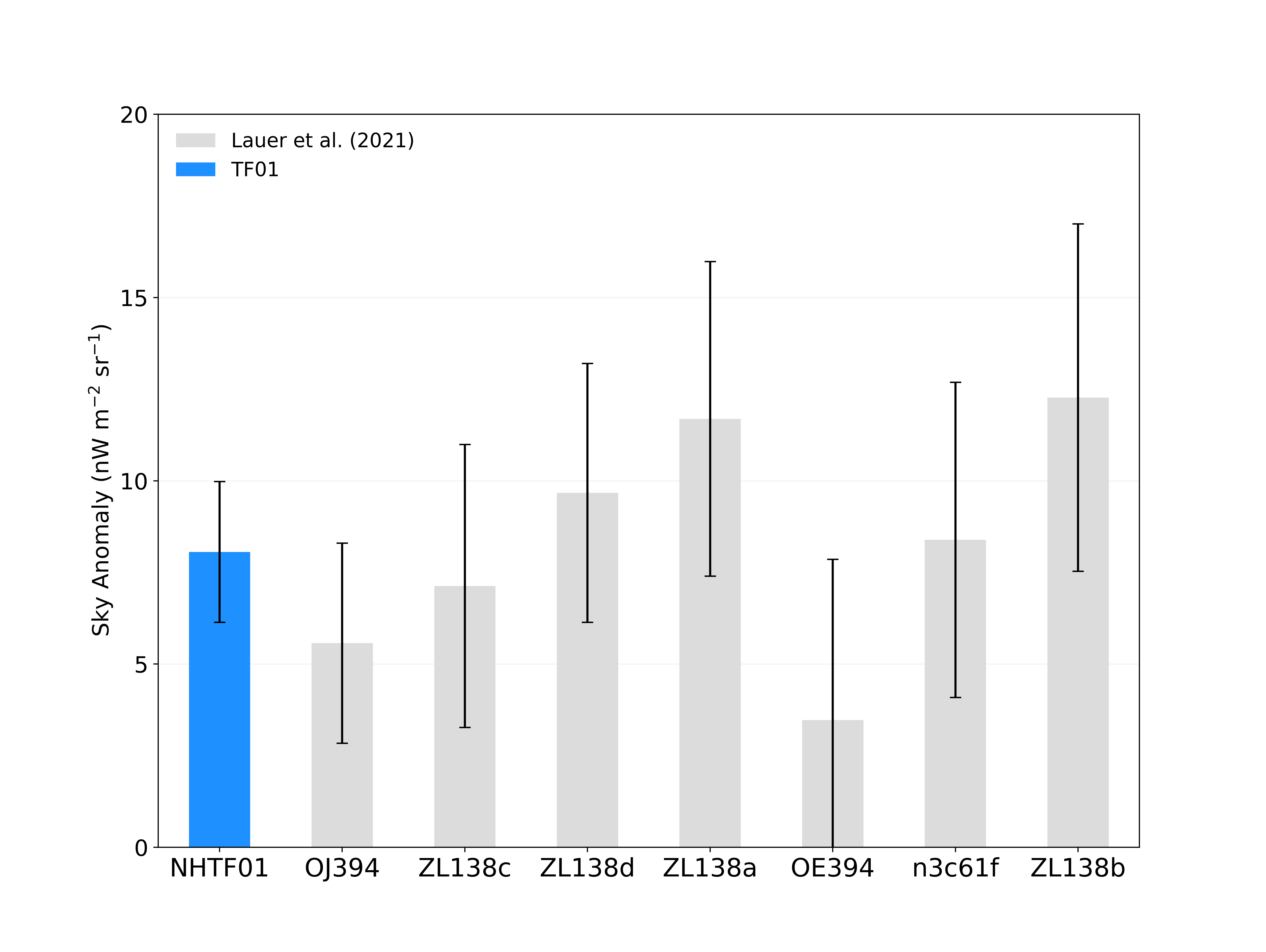}
\caption{A bar chart showing the amplitudes of the anomalous sky components for the present field (blue) as compared to the seven NH21 fields.}
\label{fig:dcobbars}
\end{figure}

We find that IGL accounts for only half of the COB. Subtracting it from the COB yields an anomalous unexplained flux component of $8.06\pm1.92~({\rm 1.16~stat.,~1.53~sys.)} {\rm  ~nW ~m^{-2} ~sr^{-1}}.$ The present anomalous sky residual as compared to those in the seven NH21 fields is shown in Figure \ref{fig:dcobbars}.  The present flux is statistically consistent with all seven previous fields, but its significance is markedly greater. 

\section{An Anomalous Background}

\subsection{The COB from Galaxy Counts and $\gamma$-ray Absorption}

A large anomalous background component would not be expected under the simple and perhaps default hypothesis that the COB and the IGL flux derived from the faint galaxies already known by HST deep counts are one and the same. The IGL has been estimated many times by many different parties, including us in NH21, and recently by \citet{driver16} and \citet{saldana}. IGL traces from both groups are plotted as a function of wavelength over the LORRI passband in Figure \ref{fig:results}. There is excellent agreement on the IGL level over the ensemble of estimates.
\citet{driver16}, \citet{saldana}, and our own estimate, all imply a contribution to the COB flux of $\sim 8{\rm  ~nW ~m^{-2} ~sr^{-1}}$ over the passband sampled by LORRI.  To be fair, these results are often based on the same observations, but this at least shows that there is little interpretive ``wiggle room" allowed in the analysis methodologies.

Very-high energy (VHE) $\gamma$-ray observations can be used to estimate the COB flux and is a completely different approach than integrating over external galaxy flux and has the virtue of depending only on the total flux density of optical photons, {\it independent of any association with a stellar system.} This is the very same quantity that we have attempted to measure with New Horizons. 
Observations of VHE ($0.1-30$ TeV) $\gamma$-rays from cosmologically distant AGN show that $\gamma$-rays are absorbed as a function of the distance of the source and the energy of the $\gamma$-ray photons \citep{hess, fermi}. Quantum electrodynamics predicts that such an effect must occur \citep{niki}.  The $\gamma$-ray photons interact with optical photons to produce $e^-/e^+$ pairs.  In effect, the ambient flux density of optical photons acts as an absorbing medium, attenuating the transmission of $\gamma$-rays over large distances.  We show the COB constraints from five recent VHE $\gamma$-ray studies: \citet{magic16}, the \citet{hess17}, the \citet{fermi}, \citet{fermi19}, and \citet{magic19} in Figure \ref{fig:results}. The concordance of the COB inferred from galaxy counts and VHE $\gamma$-ray absorption evident in Figure \ref{fig:results} is a compelling argument that the COB may well be entirely due to the light of known galaxies and holds no surprises. However, while a number of VHE $\gamma$-ray traces shown in Figure \ref{fig:results} do appear to be essentially coincident with the IGL traces, it is noteworthy that when the analysis allows for arbitrary background flux as a function of wavelength, as was done in the \citet{hess17} and \citet{magic19} papers, the VHE $\gamma$-ray constraints are markedly looser and pose no conflict with our result.

\subsection{The Actual Optical Flux Measures Imply an Anomaly}

{\it But we should be able to measure the COB flux directly with optical observations.} This is where surprises may exist. As with inferences from galaxy counts, direct detection of the COB has indeed been attempted many times by many different parties. As noted in the introduction, conducting such observations from the inner solar system is challenging, due to the strong ZL foreground.  There are many clever ways to correct for ZL, but at the penalty of large errors in the observed flux. Direct COB measures generally struggle to achieve 2-$\sigma$ detection significance of the total COB flux, let alone testing for an anomalous component. At the same time, formally, the direct flux measurements nearly always fall well above the flux implied by galaxy counts and $\gamma$-rays. 

\begin{figure}[htbp]
\centering
\includegraphics[keepaspectratio,width=3.5 in]{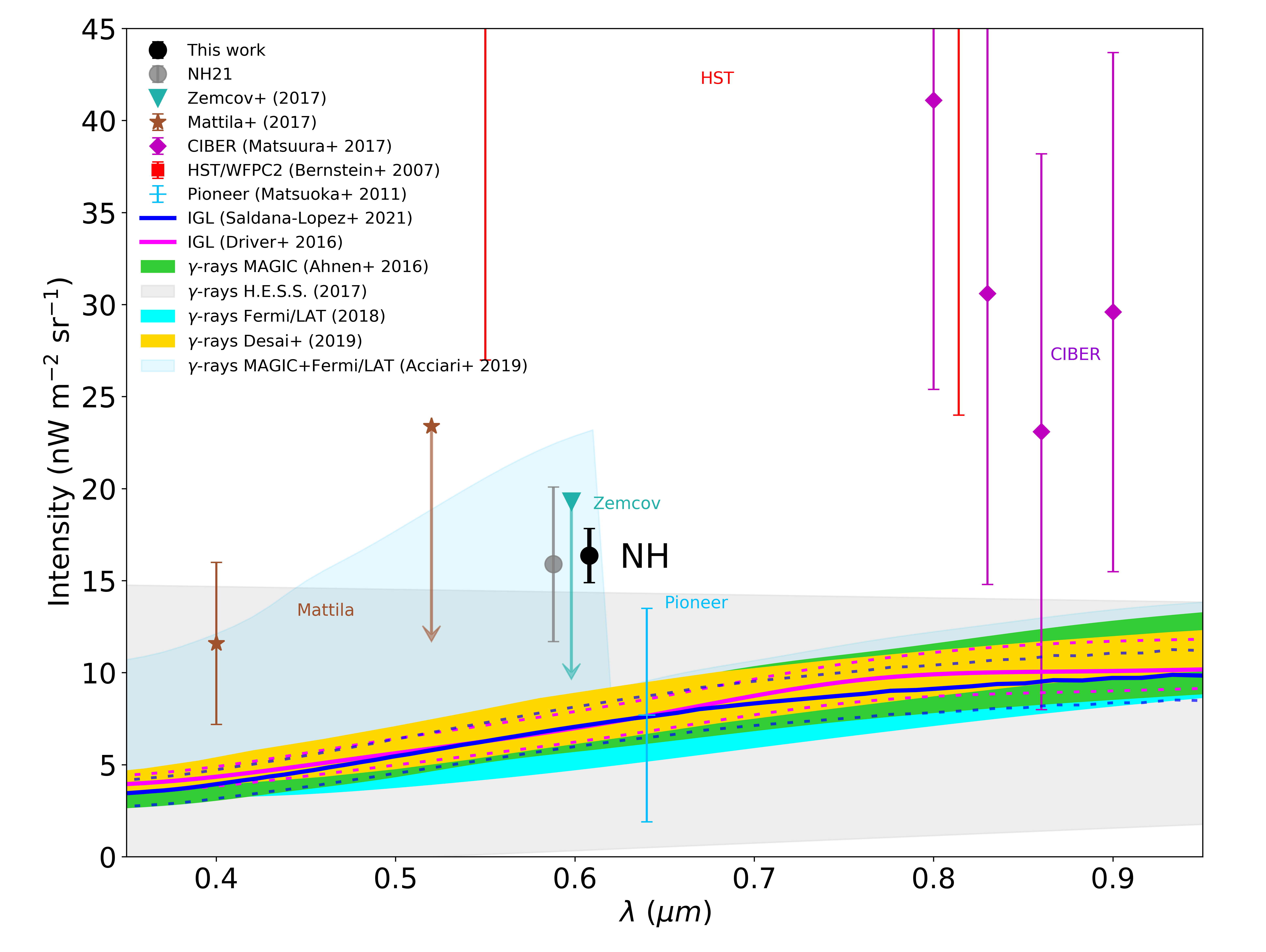}
\caption{The present result is compared to previous COB measures over the wavelengths spanned by the LORRI passband. Our NH21 COB flux (for the Zemcov DGL) is shown in gray, offset to the blue for clarity. Direct COB flux measurements are shown as points with error bars. The \citet{zemcov} flux-limit and the \citet{mattila} 0.52 ${\rm\mu m}$ limit are shown as 2-$\sigma$ upper limits with 1-$\sigma$ arrows. IGL estimates are shown as lines with 1-$\sigma$ bounds. COB fluxes inferred from VHE $\gamma$-rays are shown as shaded bands.}
\label{fig:results}
\end{figure}

Figure \ref{fig:results} shows several examples of COB measures made from Earth-space that fall within the LORRI passband. These include the HST/WFPC2 observations of \citet{wfpc2}, the CIBER rocket-based measures of \citet{ciber}, and the ``dark cloud" measures of \citet{mattila}. Of these, only the 0.40 ${\rm\mu m}$ flux of \citet{mattila} and the 0.80 ${\rm\mu m}$ CIBER flux of \citet{ciber} detect the COB with greater than 2-$\sigma$ significance. Figure \ref{fig:results} also shows the three  ``outer solar-system" COB estimates made prior to the present work. Two of these include our result from NH21, and the NH upper-limit derived by \citet{zemcov}, which we discussed in the introduction.  The third value is the COB flux derived from Pioneer 10 and 11 observations, although \citet{matsumoto} has questioned whether they are true measures of the absolute sky flux. Lastly our present COB measurement is also plotted in Figure \ref{fig:results}.  The drastically smaller error bars bracketing our result, as compared to the Earth-space measures, is due to simply having a camera far enough away from the Sun that zodiacal light doesn't matter any more.

Our COB flux is in strong tension with the integrated galaxy light flux. The implied anomalous sky component, in fact, is essentially equal to the IGL flux, itself.  The present result represents a marked improvement of over our NH21 measurement of the COB flux.  The errors bars have been reduced by a over a factor of two, greatly improving rejection of the hypothesis that the COB measured with New Horizons is consistent with the IGL. We presented a detailed discussion of this conflict in NH21, which is still valid for the present result.  In brief, \citet{conselice} has argued that the galaxy counts on which the IGL is based are strongly incomplete. \citet{cooray12}, \cite{zem14}, and \citet{mat19} have argued that the COB includes a substantial component of light from stars tidally removed from galaxies, or a population of faint sources in extended halos. None of these hypotheses may be correct, but serve to indicate that the census of extragalactic sources conducted with HST may yet be incomplete.

Finally, while many of the VHE $\gamma$-ray studies provide a constraint on the COB that is consistent with that predicted from known galaxy counts, we note speculation that the propagation of $\gamma$-rays over cosmological distances may be {partially shielded from pair-production by the VHE photons oscillating into axion-like particles (ALP) and back over their trajectory \citep{ringwald,grreview}. If this hypothesized interaction occurs,} the observed VHE $\gamma$-ray attenuation might admit COB fluxes significantly higher than the IGL flux.
\rm
\acknowledgments
We thank NASA for funding, and continued support of the New Horizons mission.  The data presented was obtained during the Kuiper Extended Mission of New Horizons. We thank Michael Coln, Steven Conard, Bruce Draine, James Gunn, James Kinnison, and David Munro for useful conversations. { We thank the referee for a prompt and thorough report, which significantly improved the paper.} This work made use of data from the European Space Agency (ESA) mission {\it Gaia} (\url{https://www.cosmos.esa.int/gaia}), processed by the {\it Gaia} Data Processing and Analysis Consortium (DPAC, \url{https://www.cosmos.esa.int/web/gaia/dpac/consortium}). Funding for the DPAC has been provided by national institutions, in particular the institutions participating in the {\it Gaia} Multilateral Agreement.
{ The Pan-STARRS1 Surveys (PS1) and the PS1 public science archive have been made possible through contributions by the Institute for Astronomy, the University of Hawaii, the Pan-STARRS Project Office, the Max-Planck Society and its participating institutes, the Max Planck Institute for Astronomy, Heidelberg and the Max Planck Institute for Extraterrestrial Physics, Garching, The Johns Hopkins University, Durham University, the University of Edinburgh, the Queen's University Belfast, the Harvard-Smithsonian Center for Astrophysics, the Las Cumbres Observatory Global Telescope Network Incorporated, the National Central University of Taiwan, the Space Telescope Science Institute, the National Aeronautics and Space Administration under Grant No. NNX08AR22G issued through the Planetary Science Division of the NASA Science Mission Directorate, the National Science Foundation Grant No. AST-1238877, the University of Maryland, Eotvos Lorand University (ELTE), the Los Alamos National Laboratory, and the Gordon and Betty Moore Foundation.}

\software{astropy \citep{2013A&A...558A..33A, 2018AJ....156..123A}, matplotlib \citep{matplotlib}, Vista \citep{vista}}

{}

\appendix

\section{Optical Photons Generated in the LORRI Optics}\label{sec:cher}

The LORRI optics include three field-flattening lenses positioned immediately in front of the CCD \citep{lorri, lorri2}. The lenses are roughly 2 cm in diameter by 0.5 cm thick, and are made of fused-silica (${\rm SiO_2}$). The LORRI CCD subtends $\sim1~{\rm sr}$ as seen from the closest element to it. A relativistic proton or electron penetrating the lenses can emit Cherenkov radiation or dislodge electrons that could excite fluorescence emission. Looking at the variety of energetic particles interacting with the lenses, it appears that $\gamma$-rays generated in the spacecraft RTG (Radioisotope Thermoelectric Generator) power supply are of the greatest concern.

While the ${\rm {}^{238}Pu}$ isotope generates the RTG power produces a low-level flux of relatively low-energy $\gamma$-rays, the trace contaminant ${\rm {}^{236}Pu}$ decays to a daughter product that generates a strong flux of 2.614 MeV photons. In 2021 the RTG is estimated to generate $2.0\times10^9$ 2.6 MeV photons s$^{-1}.$ LORRI is 2.0 m away from the nearest end of the cylindrical RTG and would receive a flux of $F_\gamma=3.9\times10^3{\rm~cm^{-2}~s^{-1}},$ assuming isotropic radiation from the RTG and no shielding. Fortuitously, LORRI is positioned only $\sim10^\circ$ off the long axis of the RTG and self absorption within the RTG is substantial.  For an RTG of New Horizons' design, the flux at this angle is $\sim5\times$ less than the isotropic assumption, or $F_\gamma=7.8\times10^2{\rm~cm^{-2}~s^{-1}},$ based on the measurements provided by \citet{jpl}.

\subsection{Cherenkov radiation from RTG $\gamma$-rays}

The 2.6 MeV  $\gamma$-rays will Compton-scatter electrons in the lenses with enough energy to produce Cherenkov radiation.  Using the Klein-Nishina equation \citep{klein}, we calculate the cross section for an ${\rm SiO_2}$ molecule to Compton-scatter a 2.6 MeV photon as $1.884\times10^{-24}$ ${\rm cm^2}.$ Given the density of fused-silica, $\rho=2.2$ ${\rm g~cm^{-3}},$ we estimate a that a lens of thickness 0.5 cm (the relevant dimension, as we argue in the next paragraph) will scatter $P_{fs}=0.021$ of the 2.6 Mev photons passing through it.

The Frank-Tamm equation \citep{jackson} provides the energy loss per unit distance traveled due to the generation of Cherenkov emission, for a relativistic electron passing through the lens. The equation gives the monochromatic energy loss at a given optical frequency and integrates it over the desired interval:
\begin{equation}
\frac{dE}{dx} =\frac{e^2}{c^2}\int_{\omega_0}^{\omega_1}\left(1-\frac{1}{\beta^2n^2(\omega)}\right)~\omega~d\omega,
\end{equation}
where $e$ is the electron charge, $\beta=v/c,$ $n(\omega)$ is the refractive index of the glass, and $\omega$ is the frequency of the light. For LORRI we are not concerned with energy loss directly, but the number of optical photons generated. Recasting the equation as the number of photons generated, using $dN=dE/(\hbar\omega),$ and taking into account that the refractive index of fused-silica ($n=1.5$) is nearly constant over the passband:
\begin{equation}
\frac{dN}{dx} =\frac{e^2}{\hbar c^2}\left(1-\frac{1}{\beta^2n^2}\right)\left[ ~\omega_1-\omega_0~\right].
\end{equation}
Compton scattering will produce electrons with a range of energies, but for typical $\beta=0.92,$ and limiting frequencies $\omega_0=2.09\times10^{15}~{\rm s^{-1}}$ and $\omega_1=4.71\times10^{15}~{\rm s^{-1}},$ corresponding to $0.9~\mu$m and $0.4~\mu$m, the Cherenkov photon production for a single scattered electron is $dN/dx =302~{\rm cm^{-1}}.$

If the RTG $\gamma$-ray flux produced isotropically-emitted Cherenkov radiation, that could account for $\sim13\%$ of the anomalous sky component.  However, the Cherenkov radiation generated by $\gamma$-rays coming from the RTG is strongly {\it anisotropic.} The following calculation of the isotropic-flux example merely serves as a point of reference to establish that the anomalous sky component cannot in fact be due to Cherenkov radiation generated in the lenses.  Potentially detectable Cherenkov-radiation photons are generated at the rate:
\begin{equation}
N_L=F_\gamma AP_{fs}~\eta_L~\frac{\Omega_L}{4\pi} ~\frac{dN}{dx}~\Delta x,
\end{equation}
where $A$ is the total area of the lenses, $\eta_L=0.9$ is the LORRI quantum efficiency, $\Omega_L=1$ sr is the solid angle of the LORRI CCD as seen from the lenses, and $\Delta x=0.3$ cm is typical length of the scattered electron's path through the lens. Each lens has area $\sim\pi$ cm$^2;$ with three lenses in series, $A=9.4$ cm$^2.$ For these parameters, $N_L=1.0\times10^{3}~{\rm s^{-1}},$ while the anomalous detected sky flux in LORRI is $8.0\times10^3$ photons s$^{-1}.$ Given the reality that Cherenkov radiation is highly anisotropic and aligned around the velocity vectors of the relativistic electrons, we will now demonstrate that all of the Cherenkov photons generated within the lenses will be directed up and out of the LORRI optics to the sky, rather than down into the CCD.  

Since, as noted, LORRI is positioned on the opposite side of the spacecraft from the RTG at an angle of only $\sim 10^\circ$ with respect to the long axis of the RTG, the RTG will appear as a relatively compact source to LORRI. Further, the $\gamma$-rays will travel outwards through LORRI roughly aligned with its optical axis.  Both Compton scattering and Cherenkov radiation have strong angular dependencies.  For Compton scattering, conservation of momentum demands that the scattered electron has a forward component of momentum aligned with the incoming $\gamma$-ray photon in addition to whatever perpendicular component is transferred to it. The trajectories of the electrons are thus confined to the hemisphere ahead of the photon. The energy imparted to the electron is a strong function of the angle of its trajectory with respect to the path of the incoming photon, with the maximum energy occurring at zero angular deflection. Conversely, electrons with large scattering angles correspond to those with low energy; for a 2.6 MeV photon, electrons deflected at angles larger than $\sim80^\circ$ will not generate Cherenkov radiation.

Cherenkov photons are emitted perpendicular to the surface of a cone with the scattered electron at its vertex; the cone's geometry is analogous to the ``Mach cone" anchored to a supersonic aircraft.  The angle of Cherenkov emission with respect to the trajectory of the electron is
\begin{equation}
\phi = \arccos{\frac{1}{\beta n}}.
\end{equation}
For LORRI, $\phi$  is always $<47\adeg3$ degrees. and the Cherenkov photons are always confined to the ``outgoing" hemisphere. The largest Cherenkev emission angle with respect to the optical axis is $87^\circ,$ which is associated with electrons scattered at $51^\circ$ from the axis (for a 2.6 MeV $\gamma$-ray moving parallel to the optical axis).  Now with the $~10^\circ$ angle of the incident photons, some small fraction of Cherenkov photons will indeed be emitted into the ``CCD hemisphere," but in directions still too far from the CCD to illuminate it.  We conclude that there is no direct path for Cherenkov photons generated by RTG $\gamma$-rays to illuminate the LORRI CCD, and thus explain the anomalous sky component.

\subsection{Fluorescent Emission induced by RTG $\gamma$-rays}

Electrons scattered by $\gamma$-rays will also lose energy by Couloumb scattering \citep{jackson} other electrons within the lenses. In the general case, as the electrons recombine with atoms within the glass, isotropically-emitted optical photons may be generated by fluorescence. Fused-silica, however, is known for its extremely low fluorescence response. a property used by \citet{moore}, for example, to allow clean isolation of Cherenkov-radiation diagnostic signals in fusion experiments.  As Moore et al.\ emphasize, ultra-pure fused silica is essentially free of optical-band fluorescent emission. Any fluorescent emission in the LORRI lenses would thus be due to trace impurities. The purity of the LORRI fused-silica glass is not known in specific detail, but the lenses were fabricated with ``standard" lens-grade material stated to have impurities at the $< 1$ ppm level.

Simple arguments based on the energetics of the RTG $\gamma$-ray flux at LORRI, as compared to the flux of the anomalous sky component, show that the anomaly is not likely to have been generated by fluorescence in the lenses.  The quantitative inputs are largely identical to those used to estimate the Cherenkov flux.  The anomalous sky signal is $8.0\times10^3$ photons s$^{-1}$ delivered to the CCD.  In a 65s exposure for an average photon energy of 2 eV (true at the $\sim6000$\AA\ pivot wavelength) the total energy received is 1.0 MeV.   The available energy provided by $\gamma$-rays is $7.8\times10^2{\rm~cm^{-2}~s^{-1}}$ 2.6 MeV photons, illuminating $9.4~\rm~cm^2$ of glass.  Only 0.021 of the photons will be scattered, and on average only 1/2 of a photon's energy will be transferred to an electron. With isotropic emission, only $(4\pi)^{-1}$ of this energy is available for generating optical photons in the CCD.  Multiplying all these factors yields a budget for generating photons of $1.1\times10^3$ MeV. As stated, pure fused-silica will absorb energy of the scattered electrons without converting it into the particular form of optical photons. With impurities in the lenses at $<10^{-6}$ abundance, even if they converted the electron energy to optical photons at 100\%\ efficiency, their net effect would be three orders of magnitude too small to account for the anomaly, if their molecular cross section for interacting with the scattered electrons is similar to that of ${\rm SiO_2}$ molecules. We conclude that fluorescent emission from the lenses is unlikely to explain the anomaly.

Lastly, we do note that the LORRI lenses are coated to reduce ghosting. Depending on the thickness of the coatings, they may effectively regarded as a separate form of impurity; two 1000\AA\ thick coatings relative to the 0.5 cm thickness of the lenses, for example, represents a $2\times 10^{-5}$ relative effect (as the coatings on each surface receive only half the electrons available to molecules in the bulk of the lenses, we have reduced their efficiency by 1/2). At this writing we have been unable to locate information on the lens coatings, but again, unless they have exceptionally large cross sections for generating optical photons, it is not likely that they can account for the anomalous sky.

\subsection{Cherenkov radiation from scattered RTG $\gamma$-rays}

At some level, structures in the New Horizons spacecraft will scatter RTG $\gamma$-rays and direct lower-energy {\it secondary} $\gamma$-rays through the LORRI optics. High-fidelity estimates of the scattered flux require a detailed structural model of the spacecraft. However, simple arguments suggest that the effects of scattered $\gamma$-rays will be modest.  The New Horizons spacecraft is optically thin to 2.6 MeV $\gamma$-rays. The fraction of $\gamma$-rays scattered is $<<1,$ and is most likely $<0.1.$  LORRI thus ``sees" the RTG surrounded by a low-amplitude $\gamma$-ray halo.  The most energetic scattered $\gamma$-rays must be those scattered only by small angles, which will also generate outwardly-directed Cherenkov photons.  Even $\gamma$-rays entering LORRI from behind at angles $45^\circ$ from the optical axis, however, will not generate Cherenkov photons that will directly illuminate the CCD.  In any case, this is where the ``isotropic" Cherenkov example is useful. If the full flux at LORRI of RTG $\gamma$-rays only produces 13\%\ of the anomalous sky component even under the (incorrect) assumption of isotropic Cherenkov radiation, a halo of scattered $\gamma$-rays down by an order of magnitude or more will certainly not be important.

\subsection{Cherenkov radiation from the RTG Neutron Flux}

The RTG also emits low-level neutron emission; however, that also appears to be insufficient by a number of orders of magnitude to generate Cherenkov radiation that would explain the anomalous sky.  As outlined by \citet{moore}, neutrons do not generate Cherenkov radiation directly, having no electric charge, but collide with the nuclei of Si and O atoms in the lenses, exciting nuclear $\gamma$-ray emission, which in turn may Compton scatter electrons. \citet{moore} provide cross sections for the generation of $\gamma$-ray sufficient to in turn generate relativistic electrons, which for a ${\rm SiO_2}$ molecule is $8.3\times10^{-27}$ ${\rm cm^2},$ over two orders of magnitude smaller than the $\gamma$-ray Compton-scattering cross section.  There is also an energy threshold; only neutrons with energies $>2$ MeV can excite the particular nuclear transitions needed to generate the relevant $\gamma$-rays.  In 2021, the neutron flux at LORRI, assuming isotropic emission from the RTG, is $89 {\rm~ cm^{-2}~s^{-1}}$ over all energies. While $\gamma$-rays will be emitted isotropically by the nuclei, the net production of $\gamma$-rays within the lenses, 0.1 s$^{-1}$ by neutrons is negligible.

\subsection{Cherenkov radiation from Cosmic Ray Protons}

Cosmic ray protons with energies $>1.34$ GeV will generate Cherenkov radiation directly in the LORRI lenses.  These will be galactic in origin and thus radiate LORRI more or less isotropically, generating isotropic Cherenkov radiation. Their flux, $1.1 {\rm~ cm^{-2}~ s^{-1}},$ at New Horizons \citep{hill} is insufficient to generate significant Cherenkov radiation.
\end{document}